# Delayed Yielding in Creep, Time - Stress Superposition and Effective Time Theory for a Soft Glass

Bharat Baldewa and Yogesh M Joshi*

Department of Chemical Engineering, Indian institute of Technology Kanpur, Kanpur 208016, India

* Corresponding author, email: joshi@iitk.ac.in

**Abstract**

In this work we investigate creep flow of aqueous suspension of Laponite, a model soft glassy material, at different aging times and stresses. We observe that this system shows time - aging time - stress superposition over a range of aging times and stresses when real time scale is transformed into effective time scale. Existence of superposition in an effective time domain facilitates prediction of long and very short time rheological behavior. Analysis of the observed behavior from effective time approach suggests that superposition is possible only when the shape of a relaxation time spectrum is preserved at various aging times and stresses. We also observe that creep curves at low aging times and greater stresses demonstrate delayed but sudden yielding. The critical time, at which material yields, increases with increase in aging time and decrease in stress. We argue that local rejuvenation of part of the glassy material causes variation in the rate of evolution of relaxation modes. The resulting interplay between aging and rejuvenating modes lead to delayed yielding as observed experimentally.

**I. Introduction**

Soft glassy materials such as concentrated suspensions and emulsions, foam, cosmetic and pharmaceutical pastes, etc. are those soft materials that are out of thermodynamic equilibrium.[1-8] In these materials, physical jamming over mesoscopic length-scales curbs the mobility of constituent elements so that material cannot explore the phase space over the experimental timescales.[2, 9] The natural tendency of any material to acquire the equilibrium state drives time dependent evolution of the microstructure to attain the progressively low free energy states as a function of time.[10, 11] This phenomenon, also known as physical aging, causes evolution of viscoelastic properties of the material as a function of time.[5, 7, 12] Application of deformation increases potential energy and induces fluidity in the material. This phenomenon is known as rejuvenation and it reverses the effect of aging. Overall the rheological behaviour of soft glassy materials is determined by interplay between aging and rejuvenation.[13, 14]

Under quiescent conditions, the relaxation dynamics of glassy materials slows down with (aging) time. Rheologically dependence of relaxation time on aging time is usually obtained by systematically carrying out creep/relaxation experiments at different aging times. The age dependent data is then shifted to obtain (process) time – aging time superposition. Struik[12] was the first to report time – aging time superposition of creep compliance of glassy polymers by plotting it against creep time divided by that factor of relaxation time which depends on aging time. However, in order to avoid effect of aging during the course of creep flow, he followed a protocol wherein he considered the creep data over only 10 % of the aging time in order to observe the time - aging time superposition. This protocol is known as Struik protocol. Creep time - aging time superposition has been reported for many amorphous polymers.[12, 15-17] Many groups have also reported time-aging time superposition for variety of soft glassy materials [18-22] (and also for spin glasses [23, 24]). Struik proposed a procedure to predict a long time creep behavior from short time tests using the superposition.[12] However, consideration of creep data over only 10 % of the aging time limited its predictive potential. Very recently Shahin and Joshi [25] proposed a formal procedure that employs effective time approach to predict long time creep behaviour from short time tests. Unlike Struik's theory, since effective time approach accounts for aging during the course of experiments, it leads to more effective prediction of long time behaviour.

Contrary to the aging phenomenon, deformation field induces greater fluidity in the material thereby reducing the relaxation time. Cloitre et al.[26] reported time – aging time superposition for soft micro-gel paste, whose relaxation time was observed to show linear dependence on aging time in the limit of small stresses. With increase in stress, dependence of relaxation time on aging time weakened. Above the yield stress material underwent complete rejuvenation eliminating dependence of relaxation time on aging time.[26] Recognizing similarity of curvature of the stress dependent data in superposition, Joshi and coworkers [27, 28] observed creep time - aging time - stress superposition for aqueous Laponite suspension in the limit of Struik protocol. Interplay between deformation field induced rejuvenation and aging has also led to demonstration of many complex rheological phenomena in soft glassy materials such as viscosity bifurcation,[29-32] sudden jamming (liquid to glass transition) under high magnitude of oscillatory stress,[33] and overaging.[34-36]

Soft glassy materials do not follow time translational invariance (TTI) owing to their time dependent behaviour.[37, 38] Because of which, the Boltzmann superposition principle[39] cannot be applied to soft glassy materials. This significantly limits the theoretical modelling of the rheological behaviour of this class of materials. In a seminal contribution Fielding and coworkers[38] used an effective time approach[40] and proposed a modified Boltzmann superposition principle. They also proposed a comprehensive theory known as soft glassy rheology (SGR) model. This model naturally accounts for physical aging through activated dynamics, by allowing strain degrees of freedom and by employing an effective temperature instead of thermodynamic temperature.[38] Remarkably SGR model captures most of the generic features of soft glassy rheological behaviour. Over past decade many groups have also proposed phenomenological models to analyse the rheological behaviour by considering interplay between aging and rejuvenation.[30, 31, 41] These models are mathematically simple and capture many of the experimental observations qualitatively.

In the present work, we investigate creep flow behaviour of a model soft glassy material: an aqueous suspension of Laponite. We observe time - aging time - stress superposition using effective time approach and demonstrate prediction of long time creep behaviour. However, over a certain range of aging times and stresses, material is observed to undergo delayed but sudden yielding in creep flow. We conclude by discussing how interaction between deformation field and relaxation time distribution may lead to various observed behaviours.

**II. Material and Experimental Procedure**

Laponite-RD,® synthetic hectorite clay was procured from Southern Clay Products Inc. Laponite RD consists of disk shaped nanoparticles having diameter around 25 to 30 nm and thickness 1 nm.[42] White powder of Laponite was dried for 4 hours at 120°C to remove moisture before mixing with ultrapure water having pH 10. Laponite was dispersed in water using Ultra Turrex drive for a period of 30 min. The suspension was then left undisturbed for 80 days in a sealed polypropylene bottle. The detailed preparation protocol to prepare aqueous Laponite suspension is discussed elsewhere.[43] In this work we have used 2.8 weight % concentration, whose suspension in water is known to form a soft solid having paste like consistency.[44]

The rheological experiments were conducted on Anton Paar MCR 501 rheometer (Couette geometry, bob diameter 5 mm and gap 0.2 mm). Couette shear cell was filled with a new sample at the beginning of each experiment and shear melted under oscillatory shear field having strain magnitude of 1500 and frequency of 0.1 Hz for 15 min. Shear melting is necessary to obtain the same initial condition in all the experiments. After the shear melting sample was left to age for a predetermined time. During the aging process we probed the evolution of elastic modulus of the suspension by applying small amplitude oscillatory shear having strain magnitude 0.005 at frequency 0.1 Hz. Subsequent to aging, a constant shear stress was applied to the suspension. In this work we have applied shear stress in the range 10 Pa to 80 Pa. In order to investigate presence of wall slip, we performed few creep experiments in a different couette cell having bob diameter 2.67 cm and gap 1.13 mm. The difference in strain associated with the two shear cells, in otherwise identical experiments, was observed to be within the experimental uncertainty. This rules out presence of any noticeable wall slip. In every experiment the free surface of suspension was covered with a thin layer of low viscosity Silicone oil to avoid contamination of $CO_2$ and evaporation of water. All the experiments were carried out at 25°C.

## III. Results and Discussion

Aqueous suspension of Laponite undergoes structural evolution as a function of time elapsed since shear melting (also known as aging time $t_w$). Figure 1 shows corresponding evolution of elastic modulus as a function of aging time. We perform creep experiments on samples having different ages (aged for different $t_w$). In an inset of figure 2, compliance induced in the material has been plotted as a function of creep time. It can be seen that lesser strain gets induced in the sample for the experiments carried out at greater aging times. This behavior is due to enhancement in both, elastic modulus as well as viscosity, as a function aging time (enhancement in elastic modulus decreases glassy compliance, while increase in viscosity makes rate of change of compliance weak at greater aging times).

In the limit of linear response, Boltzmann superposition principle is applicable to the soft materials that follow TTI and is given by:[39]

$$\gamma(t) = \int_{-\infty}^{t} J(t-t_w) \frac{d\sigma}{dt_w} dt_w, \qquad (1)$$

where $t$ is present time, $t_w$ is time at which deformation was applied to the system, $\sigma$ is shear stress imposed on the material while $\gamma$ is shear strain induced in the material. For the materials that follow TTI, compliance $J$ is only a function of time elapsed since application of deformation $(J = J(t-t_w))$ in the linear response regime. However, time dependency shown by glassy materials forbids application of the Boltzmann superposition principle to the same. In time dependent materials, creep compliance shows additional dependence on time at which creep was applied: $J = J(t-t_w, t_w)$ leading to:[37, 38]

$$\gamma(t) = \int_{-\infty}^{t} J(t-t_w, t_w) \frac{d\sigma}{dt_w} dt_w, \qquad (2)$$

For a system that shows time dependent variation of relaxation time, it is customary to transform the real time scale to the effective time scale.[40] Effective time represents the time required for an occurrence of the same relaxation processes with constant relaxation time, which otherwise occur on a real time scale ($t$) with time dependent relaxation time $\tau(t)$. The effective time associated with a constant relaxation time ($\tau_0$) is given by:[25, 37, 38]

$$\xi(t) = \int_0^t \tau_0 dt'/\tau(t') \qquad (3)$$

In this expression actual value of $\tau_0$ is immaterial for the analysis. What is important is that in the effective time domain, value of $\tau_0$ remains constant. As a result, compliance is only a function of effective time elapsed since application of creep flow field: $J = J(\xi(t) - \xi(t_w))$. Therefore the modified Boltzmann superposition principle can be written as:[25, 37, 38]

$$\gamma(t) = \int_{-\infty}^t J(\xi(t) - \xi(t_w)) \frac{d\sigma}{dt_w} dt_w \qquad (4)$$

For a glassy material, relaxation time is known to follow a power law dependence on aging time given by: $\tau = A\tau_m^{1-\mu} t_w^\mu$,[12, 25, 26, 38] where $\tau_m$ is microscopic relaxation time and $A$ is a constant. The difference in effective time is then given by:

$$\xi(t) - \xi(t_w) = \int_{t_w}^t \tau_0 dt'/\tau(t') = \frac{\tau_0 \tau_m^{\mu-1}}{A}\left[\frac{t^{1-\mu} - t_w^{1-\mu}}{1-\mu}\right] = \frac{\tau_0 \tau_m^{\mu-1} \theta(t, t_w)}{A}. \qquad (5)$$

In figure 2, we plot aging time dependent creep curves obtained at 40 Pa stress in the effective time domain. It can be seen that the vertically shifted creep curves demonstrate an excellent superposition, irrespective of their aging times. Vertical shifting, which is carried out by normalizing compliance by the modulus, is necessary to accommodate time dependent increase in elastic modulus as shown in figure 1.

In figure 3 we plot creep curves obtained at various aging times for higher creep stresses. For a creep stress of 60 Pa and at an aging time of 600 s compliance shows small strain at lower creep times. At a critical creep time, however, compliance shows a sudden enhancement. Overall the material demonstrates time delayed yielding behaviour. At greater aging times creep curves do not show delayed yielding over the explored creep times. In an inset of figure 3(a), a creep time - aging time superposition is attempted for the unyielded creep curves (creep curves associated with four higher aging times). It can be seen that creep curves associated with $t_w$ =1800, 2700, 3600 s demonstrate an excellent superposition for $\mu$ =0.8. However the creep curve associated with $t_w$ =1200 s, does not follow the superposition. Application of 65 Pa stress causes delayed yielding for creep experiments performed at higher aging times of 600, 1200 and 1800 s, as shown in figure 3 (b). Furthermore, the critical time at which material yields increases with increase in aging time. Interestingly, the creep curves that show yielding, superpose at large creep times in the real time domain. This behavior suggests complete rejuvenation and absence of aging (or absence of time dependency, $\mu$ =0). At further greater aging times, however, the compliance does not show yielding over the explored creep times. The unyielded creep curves do show superposition when normalized compliance is plotted against $\theta(t,t_w)$ for $\mu$ =0.73 as shown in an inset of figure 3 (b). Figure 3 (c) shows creep behavior at 80 Pa stress, wherein all the creep curves obtained at various aging times demonstrate delayed yielding. In the inset of figure 3 (c) we plot creep curves that show yielding, which are obtained at same age (600 s) under application of different stresses. It can be seen that irrespective of the stress, compliance follows the same curve. Overall both the inset and main plot of figure 3 (c) suggest complete rejuvenation (cessation of aging, $\mu$=0) after yielding at long creep times.

In figure 4, rate of evolution of relaxation time with respect to aging time ($\mu$) is plotted against the magnitude of stress. It can be seen that $\mu$ demonstrates bifurcation for a certain range of stresses over the explored aging times. In figure 3, three types of characteristic behaviours are reported. At lower aging times, application of stress above a certain magnitude causes delayed but sudden yielding in the material, which brings about complete rejuvenation leading to $\mu$=0. On the other hand, for higher aging times creep curves superpose leading to time - aging time superposition for a particular value of $\mu$ (>0). For an intermediate range of aging times, creep curve(s) demonstrates non-uniform rejuvenation without undergoing complete yielding, which prevents it from participating in the time - aging time superposition associated with higher aging time creep curves. The behaviour observed in figure 3 and figure 4 further suggest that even for the lower stresses (40 Pa and below), bifurcation of $\mu$ may be possible if the creep experiments are carried out at further lower aging times. Similarly at greater magnitude of stresses, creep curves at very high aging times may demonstrate superposition for nonzero values of $\mu$.

In the inset of figure 4, we plot time - aging time superpositions obtained at different stresses that require nonzero $\mu$. It can be seen that the respective stress dependent superpositions have similar curvatures. The horizontal and vertical shifting of these superpositions on to a superposition associated with 40 Pa stress, which we plot in figure 5, produces time - aging time - stress superposition. In this superposition we use the creep data belonging to 40 Pa stress only up to creep time of 100 s. This is carried out in order to facilitate prediction of the remaining data as discussed below. It should be noted that the self-similarity of curvature leading to the superposition is not necessarily a proof of the physical existence of the same and therefore needs further validation. The time - aging time superpositions at different creep stresses shown in the inset of figure 4 have been plotted in the effective time domain with abscissa being $\left[\xi(t)-\xi(t_w)\right]/\left[\tau_0 \tau_m^{\mu-1}/A\right]$. In the effective time domain, relaxation time of the material is constant and according to equation 3, we consider it to be $\tau_0$. Value of $\mu$ associated with every stress dependent superposition shown in the inset of figure 4 is different. Therefore, in order to obtain time – aging time – stress superposition (by representing the abscissa by $\left[\xi(t)-\xi(t_w)\right]/\tau_0$), horizontal shift factor $(c)$ in figure 5 must scale as $\tau_m^{\mu-1}/A$. Since all the stress dependent superpositions have been shifted on to a superposition belonging to reference stress (40 Pa), horizontal shift factor should be: $c = \tau_m^{\mu-1}/\tau_m^{\mu_R-1}$, where $\mu_R$ is value of $\mu$ associated with reference superposition. In the inset of figure 5 we plot $\ln c$ as a function of $\mu-1$. Observed linear relationship between both the variables ($c \propto \tau_m^{\mu-1}$) indeed validates the physical existence of the superposition.

The time - aging time - stress superposition shown in figure 5 can be used to predict the long and short time creep data. Since in an effective time domain relaxation time is constant, the superposition shown in figure 5 represents creep behavior of a material with modulus equal to that associated with reference aging time (3600 s) under application of reference shear stress (40 Pa) plotted against creep time (in effective time domain) normalized with (constant) relaxation time. The real time creep behavior can be obtained by transforming the superposition shown in figure 6 from an effective time domain to real time domain. For $\mu=\mu_R$, abscissa of figure 5 is represented by $\theta(t,t_w)$, which can be transformed to real time domain by inverting equation (5) to give:

$$t-t_w = \left\{\theta(1-\mu)+t_w^{1-\mu}\right\}^{1/(1-\mu)} - t_w. \tag{6}$$

In figure 6 we plot compliance predicted from the superposition shown in figure 5 as a function of creep time $t-t_w$ using equation 6. The experimental data for which prediction is proposed is same as that given in the inset of figure 2 ($\sigma$ = 40 Pa, $t_w$ =600 to 3600 s). The filled symbols represent that creep data which is part of the superposition, while open symbols represent that data which is not. The lines represent prediction of the creep behavior. It can be seen that the lines not only show an excellent fit to the data, which is not a part of the superposition; but also show prediction of the long time and short time creep behavior for which data is not available. On one side prediction of long term creep behavior has its own importance, while on the other side prediction of short time creep data is also equally important. This is because instrument inertia forbids data acquisition at very small timescales (as can be clearly observed from the initial oscillations in the creep data). In previous work, we proposed a methodology to predict long and short time creep behavior from time - aging time superposition.[25] The prediction using time - aging time - stress superposition clearly gives an advantage; for example, time- aging time superposition shown in figure 2 makes available the information of creep behavior in the range: 0.01 to 1.8 in scale of $\theta(t,t_w)$ for stress of 40 Pa. However, for time - aging time - stress superposition the information is available in the range: 0.009 to 10 for the same stress. According to equation 6, time - aging time - stress superposition therefore offers significant advantage in predicting very long and short time rheological behavior compared to time - aging time superposition.

Although the predictive capacity demonstrated in figure 6 appears very promising, behavior reported in figure 3 raises an important question associated with validity of such prediction over the range of aging times and stresses. Therefore, it is necessary to discuss precautions that need to be exercised before employing time - aging time - stress superposition to predict the long time and short time creep behavior. Furthermore, Struik[12] suggested that time - aging time superposition is possible if aging affects only the average value of relaxation times and not a shape of the spectrum. Below we also address this issue from a point of view of effective time theory.

In soft glassy materials, entities (particles) that constituent the same are arrested in physical cages formed by their neighbors, which are generically represented by energy wells.[45] Typically there exists a distribution of energy well depths (or the barrier heights) in which entities are arrested. The distribution of energy well depths also implies a spectrum of relaxation times through its proposed Arrhenius dependence on energy well depth given by: $\tau_i = \tau_m \exp(E_i/kT)$.[38] By virtue of system being arrested in a jammed state, the depths of majority of energy wells are greater than the thermal energy associated with the entities. In an aging process, constituents undergo structural rearrangement and decrease their energy (increase well depth) as a function of time. This process causes progressive enhancement in relaxation times.[10, 12] In figure 7 we represent this scenario by a schematic, wherein a relaxation time distribution is represented at three aging times, wherein shape of the spectrum is shown to be unaffected by the aging process. For a material with single relaxation mode, as shown by equations (2) and (4), time elapsed since application of deformation is needed to be replaced by effective time elapsed since application of deformation. However, if the material possesses many relaxation times described by a spectrum, then there will be a new effective time scale associated with every relaxation mode in the spectrum. Let us assume that the material possesses $n$ relaxation modes represented by: $\tau_i$ ($i$ = 1 to $n$). Let each relaxation mode evolve with aging time according to: $\tau_i = A_i \tau_m^{1-\mu_i} t_w^{\mu_i}$, where $\mu_i$ is logarithmic rate of aging ($d\ln\tau_i / d\ln t_w$) associated with $i$ th mode. The corresponding effective time scale is then given by:

$$\xi_i(t) - \xi_i(t_w) = \int_{t_w}^{t} \tau_0 dt' / \tau_i(t') = \frac{\tau_0 \tau_m^{\mu_i - 1}}{A_i} \left[ \frac{t^{1-\mu_i} - t_w^{1-\mu_i}}{1-\mu_i} \right]. \tag{7}$$

Creep compliance will, therefore, depend on effective time scale associated with each mode:

$$J = J\left( [\xi_1(t) - \xi_1(t_w)], \ldots, [\xi_i(t) - \xi_i(t_w)], \ldots, [\xi_n(t) - \xi_n(t_w)] \right) \tag{8}$$

If, however, every relaxation mode demonstrates the same dependence on aging time, such that: $\mu_i = \mu$ $\{i = 1 \text{ to } n\}$, the difference in effective time for $i$ th mode will be:

$$\xi_i(t) - \xi_i(t_w) = \frac{\tau_0 \tau_m^{\mu-1}}{A_i} \left[ \frac{t^{1-\mu} - t_w^{1-\mu}}{1-\mu} \right], \tag{9}$$

and if $\tau_a \{= A_a \tau_m^{1-\mu} t_w^{\mu}\}$ is average relaxation time, we get:

$$\xi_i(t) - \xi_i(t_w) = \frac{A_a}{A_i} \left( \xi_a(t) - \xi_a(t_w) \right), \tag{10}$$

where, subscript $a$ represents variables associated with the average relaxation time. Finally equations (8) and (10) lead to:

$$J = J\left(\left[\frac{A_a}{A_1}\left(\xi_a(t) - \xi_a(t_w)\right)\right], \ldots, \left[\frac{A_a}{A_i}\left(\xi_a(t) - \xi_a(t_w)\right)\right], \ldots, \left[\frac{A_a}{A_n}\left(\xi_a(t) - \xi_a(t_w)\right)\right]\right).$$

(11)

Since, $A_i$ are constants, compliance can be represented as a function of difference in effective times associated with only average relaxation mode:

$$J = J\left(\xi_a(t) - \xi_a(t_w)\right),$$  (12)

which according to equations (4) and (5) is a sufficient requirement to observe the superposition. Therefore, according to effective time theory, if all the relaxation modes evolve with the same value of $\mu$, then system demonstrates time - aging time superposition.

Application of deformation field (in the present case, stress field) displaces the constituents originally arrested in their own cages with respect to each other. Consequently, the strain ($\gamma$), which is assumed to get induced in the material affinely, enhances potential energy of the constituents trapped in their (quadratic) energy wells by $k\gamma^2/2$, where $k$ is an elastic spring constant.[38] We have represented this scenario by a schematic in figure 8, wherein proposed situation arising from application of stresses of different magnitude on a soft glassy material having same age is represented. It can be seen that strain induced in the material enhances the potential energy of a trapped constituent. This reduces the energy barrier required to escape the well and is given by: $\Delta E_i = E_i - \frac{1}{2}k\gamma^2$. Consequently escape time (characteristic relaxation time) also reduces according to:[38] $\tau_i = \tau_m \exp(\Delta E_i / k_B T)$. Logarithmic rate of change of $i$ th relaxation mode is then given by:

$$\mu_i = d\ln\tau_i / d\ln t_w = \frac{1}{k_B T} d\left(E_i - \frac{k\gamma^2}{2}\right) \Big/ d\ln t_w. \tag{13}$$

In the inset of figure 4, we observe that every time- aging time superposition is characterized by a constant value of $\mu$. Therefore, if a given magnitude of stress affects all the relaxation modes in same fashion such that associated values of $\mu_i$ for a given stress are same for all the modes, then time - aging time superposition will be possible at that stress.

In principle, a greater magnitude of stress at any given time will induce greater magnitude of strain in a material having same age. Consequently, the distribution of energy barrier will shift towards the lower energies as shown in schematic represented in figure 8. However, if the strain induced in the material is not significant so that $k\gamma^2/2$ is smaller than the energy well depth of the particles trapped in shallow wells, application of stress field will not alter the shape of a distribution of relaxation times, but only the average value (It is known that viscosity of a material depends more strongly on slower relaxation modes;[39] therefore even if very fast modes do get affected by the deformation field, rheological behavior will still be determined by a shape of the spectrum associated with moderately fast to slow modes). Therefore, if a material follows time - aging time superposition in creep (by fulfilling condition illustrated in figure 7), it will also follow time - aging time - stress superposition for those stresses that do not significantly affected a shape of the spectrum (by fulfilling condition illustrated in figure 8). Figure 4 suggests that value of $\mu$ decreases with increase in stress. Equation 13, therefore, suggests that intensity of physical aging, which is represented by rate of change of $E_i - \frac{k\gamma^2}{2}$ as a function of $\ln t_w$, becomes weaker with increase in stress. Furthermore, similar to that observed for a single mode analysis, even though $\mu$ decreases with stress, since shape of a spectrum is preserved, the system will demonstrate time - aging time - stress superposition over the range of stresses. Therefore, effective time theory clearly suggest that prediction of long or short time behavior is possible only if a shape of the spectrum remains unaltered for the duration over which the prediction is sought.

If stress applied to the system is large, strain induced in the material will enhance potential energy of the particles to a greater extent causing local yielding of significant fraction of the particles. However those particles that are trapped in deeper wells shall continue to undergo aging and increase their energy barrier (decrease the potential energy) as a function of time. The whole dynamics of the material in such state will be determined by interplay between aging of particles in deep wells and rejuvenation of particles in shallow wells. Let us consider an aging system at different ages. By virtue of greater average barrier height for an older systems as shown in figure 7, their average relaxation time and viscosity will also be greater compared to the younger systems. Therefore, application of stress having same magnitude will induce greater strain in the younger samples at any given creep time. This will induce rejuvenation of a greater fraction of particles in younger samples than in older samples. Consequently, the smaller fraction of particles will continue to undergo aging dynamics in younger samples compared to that of in older samples. In addition, if the rate of enhancement of strain is such that the term $k\gamma^2/2$ dominates the overall enhancement in barrier height as a function of aging time, progressively greater fraction of particles will undergo rejuvenation as a function of time. As a result, this process will further reduce fraction of particles that are aging and eventually, by a forward feedback mechanism, a sudden yielding will occur as shown in figure 3. On the other hand, if aging dynamics of the particles trapped in deeper wells is strong enough so that progressive increase in strain is not able to enhance fraction of particles getting rejuvenated any further, sudden yielding will not be observed. However in the latter case, the rejuvenation dynamics will change a shape of the spectrum of relaxation times. Therefore, such creep curve will not participate in time - aging time superposition as demonstrated in the inset of figure 3a (creep curve belonging to $t_w$ =1200 s).

**IV. Conclusion:**

In this work we study creep behavior of aging aqueous suspension of Laponite at various aging times (time elapsed since mechanical quench) and stresses. We use effective time approach to analyze the creep data. This approach facilitates transformation of real time scale (with time dependent relaxation modes) to effective time scale (with constant relaxation mode) so that Boltzmann superposition principle is applicable. We observe that the creep curves obtained over a range of aging times for a given creep stress superpose to demonstrate creep time - aging time superposition when plotted in effective time scale. Such superposition also leads to estimation of the rate of evolution of relaxation time as a function of aging time ($\mu$). Time - aging time superpositions obtained at different stresses, which lead to decrease in $\mu$ with increase in stress, produce time - aging time - stress superposition with appropriate modification of effective time scale. Existence of such superposition in effective time domain facilitates prediction of long and very short time rheological behavior. We observe that time - aging time - stress superposition demonstrates greater predictive capacity compared to that from time - aging time superposition at a single stress. Analysis of the observed behavior from a point of view of effective time theory suggests that time - aging time - stress superposition is possible only when a shape of the spectrum of relaxation times is preserved as a function of aging time and applied stress.

The creep curves obtained at small aging times and very high stresses do not participate in the superposition and are observed to undergo delayed but sudden yielding. The critical creep time at the onset of yielding is observed to increase with increase in aging time and decrease in stress. We propose that rejuvenation of part of the relaxation modes under application of strong deformation field leads to nonlinear coupling between modes that are aging and modes that are undergoing rejuvenation. Consequently progressive rejuvenation of increasingly greater fraction of relaxation modes lead to delayed yielding as observed experimentally.

**Acknowledgement:** This work was supported by Department of Science and Technology, Government of India under IRHPA scheme.

**References**

1. N. Koumakis and G. Petekidis, *Soft Matter*, 2011, **7**, 2456-2470.
2. L. Cipelletti and L. Ramos, *J. Phys. Cond. Mat.*, 2005, **17**, R253–R285.
3. P. Coussot, *Lecture Notes in Physics*, 2006, **688**, 69-90.
4. M. E. Cates and M. R. Evans, eds., *Soft and fragile matter*, The institute of physics publishing, London, 2000.
5. G. B. McKenna, T. Narita and F. Lequeux, *Journal of Rheology*, 2009, **53**, 489-516.
6. S. A. Rogers, P. T. Callaghan, G. Petekidis and D. Vlassopoulos, *Journal of Rheology*, 2010, **54**, 133-158.
7. B. M. Erwin, D. Vlassopoulos, M. Gauthier and M. Cloitre, *Physical Review E*, 2011, **83**, 061402.
8. X. Di, K. Z. Win, G. B. McKenna, T. Narita, F. Lequeux, S. R. Pullela and Z. Cheng, *Physical Review Letters*, 2011, **106**, 095701.
9. A. J. Liu and S. R. Nagel, *Nature*, 1998, **396**, 21-22.
10. D. J. Wales, *Energy Landscapes*, Cambridge University Press, Cambridge, 2003.
11. H. B. Callen, *Thermodynamics and an introduction to thermostatistics*, John Wiley & Sons, New York, 1985.
12. L. C. E. Struik, *Physical Aging in Amorphous Polymers and Other Materials*, Elsevier, Houston, 1978.
13. F. Ozon, T. Narita, A. Knaebel, G. Debregeas, P. Hebraud and J. P. Munch, *Physical Review E*, 2003, **68**, 324011-324014.
14. F. Ianni, R. Di Leonardo, S. Gentilini and G. Ruocco, *Phys. Rev. E*, 2007, **75**, 011408.
15. P. A. O'Connell and G. B. McKenna, *Polym. Eng. Sci.*, 1997, **37**, 1485-1495.
16. P. A. O'Connell and G. B. McKenna, *Mechanics Time-Dependent Materials*, 2002, **6**, 207-229.

17. S. L. Simon, D. J. Plazek, J. W. Sobieski and E. T. McGregor, *Journal of Polymer Science, Part B: Polymer Physics*, 1997, **35**, 929-936.
18. V. Awasthi and Y. M. Joshi, *Soft Matter*, 2009, **5**, 4991–4996.
19. G. R. K. Reddy and Y. M. Joshi, *Journal of Applied Physics*, 2008, **104**, 094901.
20. C. Derec, A. Ajdari, G. Ducouret and F. Lequeux, *C. R. Acad. Sci., Ser. IV Phys. Astrophys.*, 2000, **1**, 1115-1119.
21. C. Derec, G. Ducouret, A. Ajdari and F. Lequeux, *Physical Review E*, 2003, **67**, 061403.
22. P. Coussot, H. Tabuteau, X. Chateau, L. Tocquer and G. Ovarlez, *J. Rheol.*, 2006, **50**, 975-994.
23. G. F. Rodriguez, G. G. Kenning and R. Orbach, *Physical Review Letters*, 2003, **91**, 037203.
24. P. Sibani and G. G. Kenning, *Physical Review E*, 2010, **81**, 011108
25. A. Shahin and Y. M. Joshi, *Phys. Rev. Lett.*, 2011, **106**, 038302.
26. M. Cloitre, R. Borrega and L. Leibler, *Phys. Rev. Lett.*, 2000, **85**, 4819-4822.
27. Y. M. Joshi and G. R. K. Reddy, *Phys. Rev. E*, 2008, **77**, 021501-021504.
28. A. Shaukat, A. Sharma and Y. M. Joshi, *Rheologica Acta*, 2010, **49**, 1093-1101.
29. S. M. Fielding, M. E. Cates and P. Sollich, *Soft Matter*, 2009, **5**, 2378-2382.
30. P. Coussot, Q. D. Nguyen, H. T. Huynh and D. Bonn, *Phys. Rev. Lett.*, 2002, **88**, 1755011-1755014.
31. P. Coussot, Q. D. Nguyen, H. T. Huynh and D. Bonn, *Journal of Rheology*, 2002, **46**, 573-589.
32. A. Negi and C. Osuji, *Europhysics Letters*, 2010, **90**, 28003.
33. A. Shukla and Y. M. Joshi, *Chemical Engineering Science*, 2009, **64**, 4668 -- 4674
34. V. Viasnoff, S. Jurine and F. Lequeux, *Faraday Discussions*, 2003, **123**, 253-266.
35. V. Viasnoff and F. Lequeux, *Phys. Rev. Lett.*, 2002, **89**, 065701.
36. R. Bandyopadhyay, H. Mohan and Y. M. Joshi, *Soft Matter*, 2010, **6**, 1462-1468.
37. S. M. Fielding, Ph.D. Thesis, University of Edinburgh, 2000.
38. S. M. Fielding, P. Sollich and M. E. Cates, *Journal of Rheology*, 2000, **44**, 323-369.
39. J. D. Ferry, *Viscoelastic Properties of Polymers*, John Wiley, New York, 1980.
40. I. L. Hopkins, *J. Polym. Sci.*, 1958, **28**, 631-633.
41. C. Derec, A. Ajdari and F. Lequeux, *European Physical Journal E*, 2001, **4**, 355-361.
42. http://www.laponite.com.
43. A. Shahin and Y. M. Joshi, *Langmuir*, 2010, **26**, 4219–4225.
44. Y. M. Joshi, *J. Chem. Phys.*, 2007, **127**, 081102.
45. J. P. Bouchaud, *J. Phys. I*, 1992, **2**, 1705-1713.

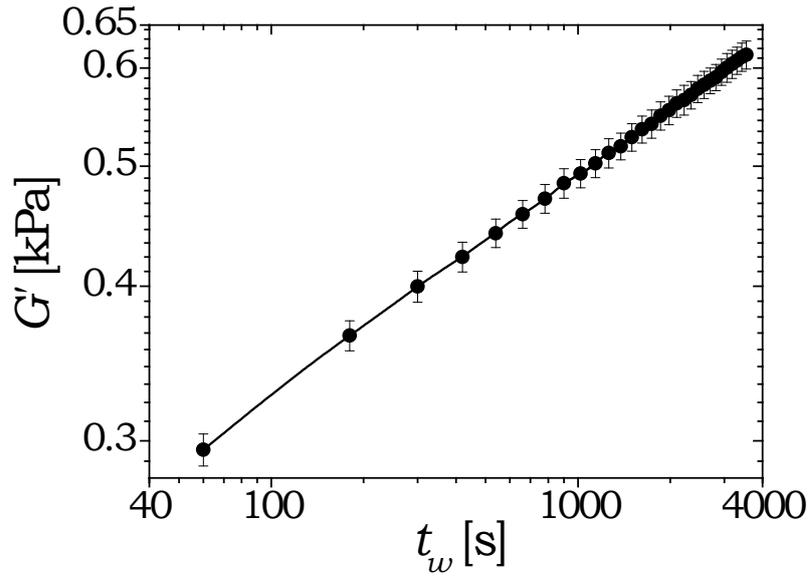

**Figure 1.** Evolution of elastic modulus ($G'$) as a function of aging time ($t_w$).

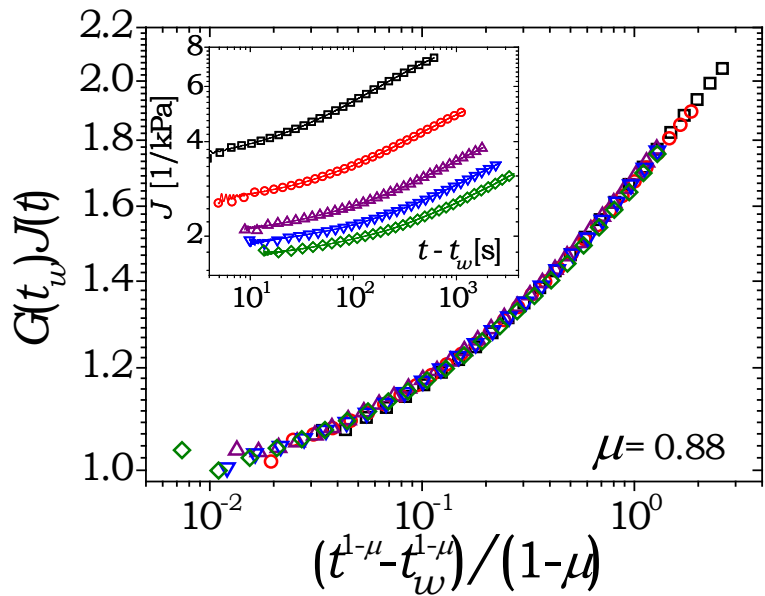

**Figure 2.** Time - aging time superposition of the creep data obtained at stress of 40 Pa using effective time approach. The inset shows compliance induced in the material as a function of creep time for creep experiments performed at various aging times, from top to bottom: $t_w$ = 600, 1200, 1800, 2700, 3600 s.

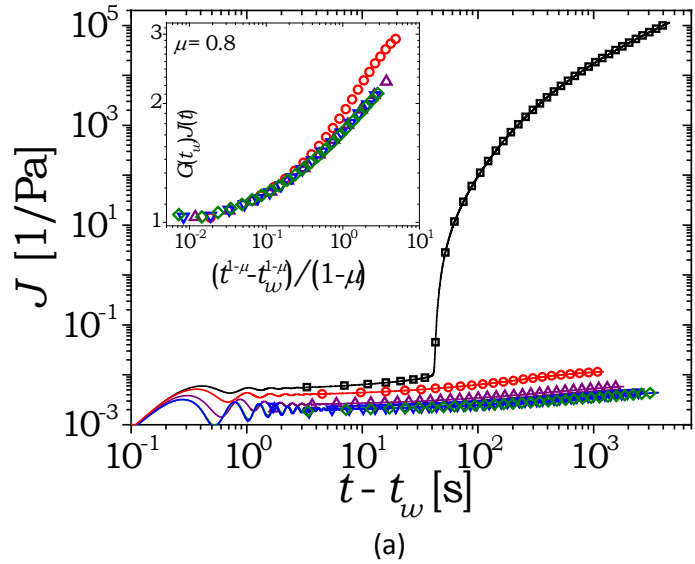

(a)

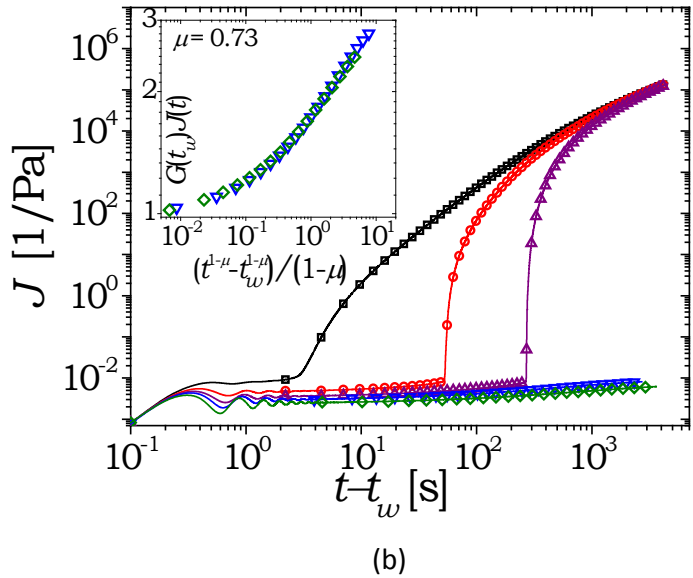

(b)

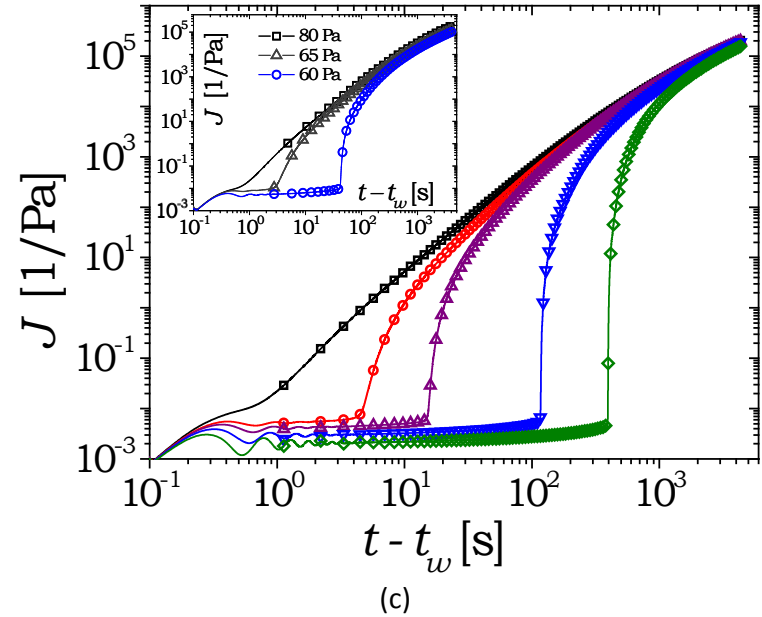

(c)

**Figure 3.** Evolution of compliance is plotted as a function of creep time for creep experiments carried out at different aging times and various stresses: (a) 60 Pa, (b) 65 Pa and (c) 80 Pa. In all the figures from top to bottom: $t_w$ = 600, 1200, 1800, 2700, 3600 s. The insets in figure (a) and (b) represent corresponding time - aging time superposition of unyielded creep curves obtained from effective time approach. The inset in figure (c) represents evolution of compliance plotted against creep time for which delayed yielding is observed, for experiments carried out at different stresses but at the same aging time $t_w$ =600 s.

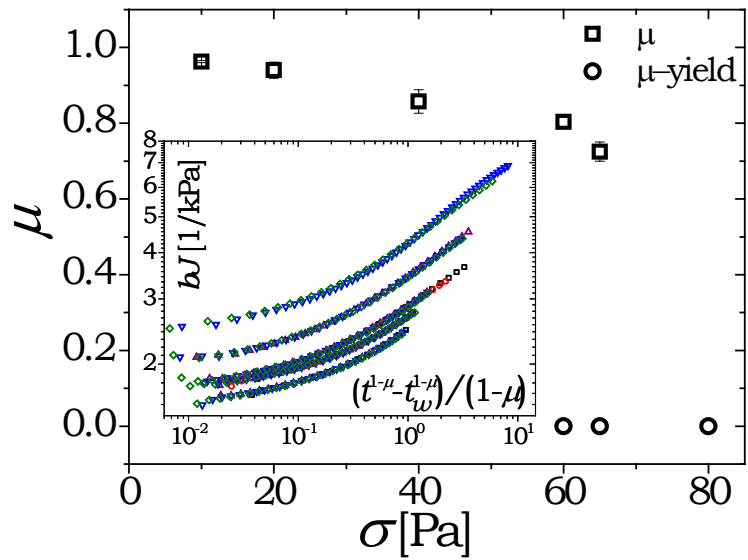

**Figure 4.** Factor $\mu$ plotted as a function of shear stress. It can be seen that $\mu$ bifurcates due yielding observed at low aging times. The inset shows individual time aging time superpositions at different stresses, from top to bottom 65, 60, 40, 20, and 10 Pa. For 60 and 65 Pa we have considered only those creep curves that participate in superposition. The vertical shift factor is given by $b = G(t_w)/G(t_{wR})$, where $G(t_{wR})$ is elastic modulus at the reference aging time ($t_{wR}$=3600 s).

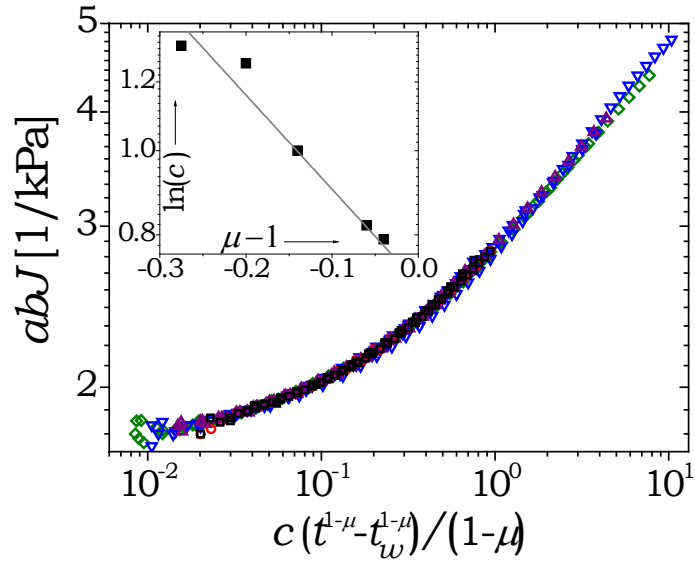

**Figure 5**. Time - aging time - stress superposition obtained by horizontally and vertically shifting individual time - aging time superpositions shown in the inset of figure 4. For 40 Pa stress we have included creep curves only up to creep time of 100 s (refer to text for details). Vertical shift factor is given by $a = G(t_{wR}, \sigma)/G(t_{wR}, \sigma_R)$, where $\sigma_R = 40$ Pa is the reference shear stress. In the inset horizontal shift factor $c\left\{= \tau_m^{\mu-1}/\tau_m^{\mu_R-1}\right\}$ is plotted as a function of $\mu - 1$.

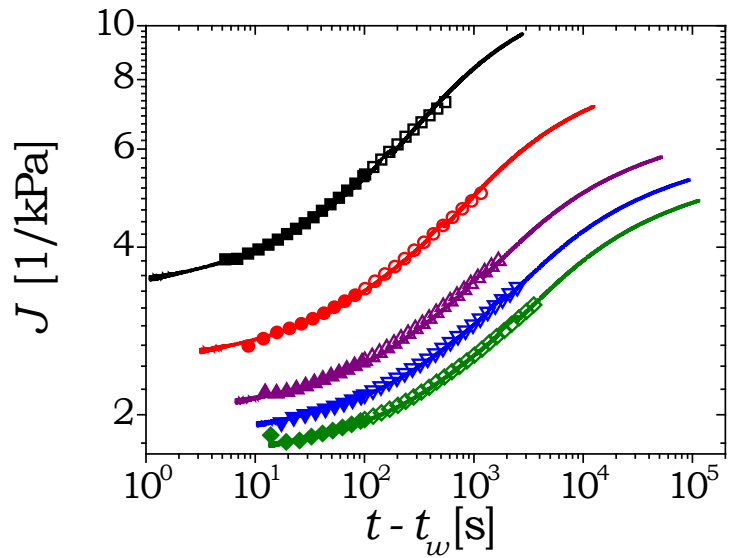

**Figure 6.** Prediction of very long time and very short time creep behavior from time - aging time - stress superposition shown in figure 5. The symbols represent creep data shown in the inset of figure 2, while the lines through respective data set represent prediction.

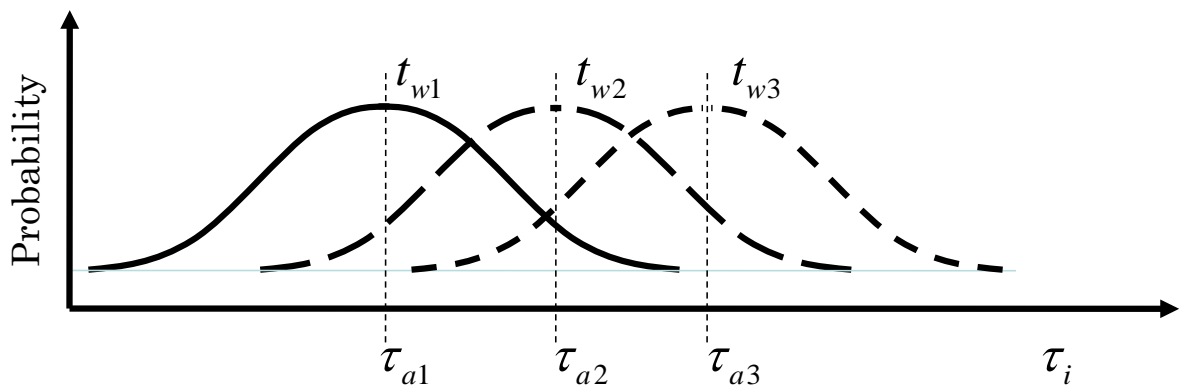

**Figure 7.** A schematic representing evolution of relaxation time distribution for various aging times ($t_{w1} < t_{w2} < t_{w3}$). Independence of a shape of the spectrum on aging time is a necessary condition to observe time - aging time superposition.

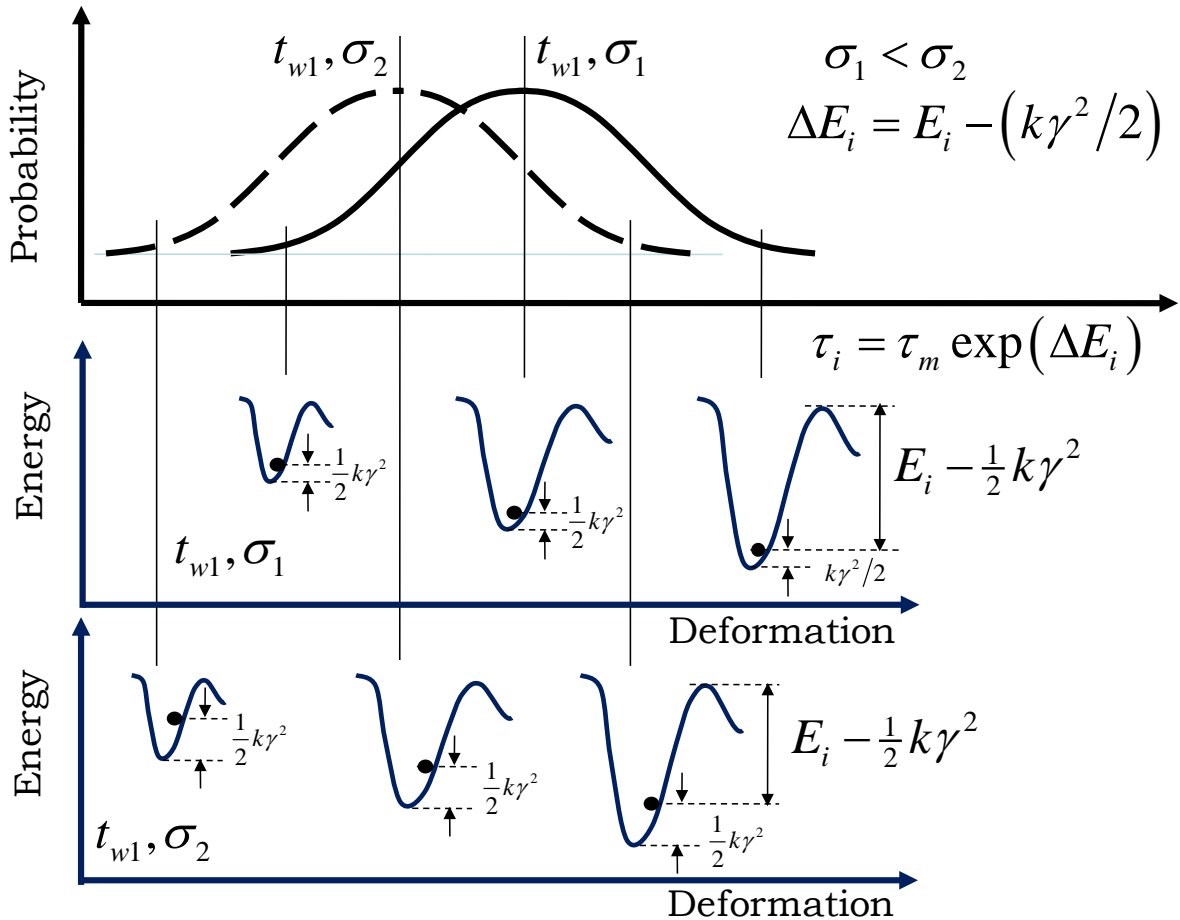

**Figure 8.** A schematic representing effect of stress on relaxation time distribution having same age ($t_{w1}$). Relaxation time is assumed to depend on energy barrier and strain according to: $\tau_i = \tau_m \exp\left(\left(E_i - k\gamma^2/2\right)/k_B T\right)$.[37, 38] Depending on strain induced in the material at two stresses ($\sigma_1 < \sigma_2$), energy barrier decreases. However, if strain induced in the material is such that that potential energy enhancement $k\gamma^2/2$ is significantly smaller than energy well depth of shallow wells in the distribution, shape of the spectrum is not affected by the deformation field.